\documentstyle{article}

\def\greaterthansquiggle{\raise.3ex\hbox{$>$\kern-.75em\lower1ex\hbox{$\sim$}}}
\def\lessthansquiggle{\raise.3ex\hbox{$<$\kern-.75em\lower1ex\hbox{$\sim$}}}
\newcommand{\beq}{\begin{equation}}
\newcommand{\beqn}{$$}
\newcommand{\eeqn}{$$}
\newcommand{\eeq}{\end{equation}}
\newcommand{\beqa}{\begin{eqnarray}}
\newcommand{\eeqa}{\end{eqnarray}}
\newcommand{\beqan}{\begin{eqnarray*}}
\newcommand{\eeqan}{\end{eqnarray*}}
\newcommand{\ba}{\begin{array}}
\newcommand{\ea}{\end{array}}

\def\nz{\ifmmode {I\hskip -3pt N} \else {\hbox {$I\hskip -3pt N$}}\fi}
\def\zz{\ifmmode {Z\hskip -4.8pt Z} \else
       {\hbox {$Z\hskip -4.8pt Z$}}\fi}
\def\qz{\ifmmode {Q\hskip -5.0pt\vrule height6.0pt depth 0pt
       \hskip 6pt} \else {\hbox
       {$Q\hskip -5.0pt\vrule height6.0pt depth 0pt\hskip 6pt$}}\fi}
\def\rz{\ifmmode {I\hskip -3pt R} \else {\hbox {$I\hskip -3pt R$}}\fi}
\def\cz{\ifmmode {C\hskip -4.8pt\vrule height5.8pt\hskip 6.3pt} \else
       {\hbox {$C\hskip -4.8pt\vrule height5.8pt\hskip 6.3pt$}}\fi}
\def\au{{\setbox0=\hbox{\lower1.36775ex%
\hbox{''}\kern-.05em}\dp0=.36775ex\hskip0pt\box0}}
\def\ao{{}\kern-.10em\hbox{``}}

\newtheorem{Theorem} {Theorem} [section]

          \newcommand{\R}{{\bf R}}

{\catcode `\@=11 \global\let\AddToReset=\@addtoreset}
\AddToReset{equation}{section}

\newcommand{\subjclass}[1]{}

\newcommand{\bysame}{---}

\def\scri{\hbox{${\cal J}$\kern -.645em {\raise
      .57ex\hbox{$\scriptscriptstyle (\ $}}}}

 \newcommand{\eq}[1]{(\ref{#1})}

\newcommand{\commentout}[1]{}

\newcommand{\ee}{\end{equation}} \newcommand{\bea}{\begin{eqnarray}}
\newcommand{\eea}{\end{eqnarray}}
\newcommand{\beaa}{\begin{eqnarray*}}
\newcommand{\eeaa}{\end{eqnarray*}}

\newcommand{\doc}{\ll\scri\gg}
\newcommand{\ext}{\mbox{\footnotesize ext}}
\newcommand{\Sext}{\Sigma_{\mbox{\footnotesize ext}}}
\newcommand{\Sint}{\Sigma_{\mbox{\footnotesize int}}}
\newcommand{\sing}{{\cal S}}
\begin{document}

\title{All electro--vacuum
Majumdar--Papapetrou space--times with nonsingular black holes}
\author{ Piotr T.\ Chru\'sciel\thanks{
 On leave of absence from the Institute of Mathematics, Polish Academy
    of Sciences, Warsaw.  Supported in part by KBN grant \# 2 1047
    9101  and by the Federal Ministry of
    Science and Research, Austria.  {\em E--mail}:
    Chrusciel@Univ-Tours.fr}
\\ D\'epartement de Math\'ematiques\\Facult\'e
des Sciences\\ Parc de Grandmont\\ F37200 Tours, France
\\ \\
Nikolai S.\ Nadirashvili\thanks{ {\em E--mail}: nnadiras@keen.esi.ac.at}
 \\ E. Schr\"odinger Institute \\
Pasteurgasse 6/7 \\ A 1090 Wien, Austria}

\maketitle

\begin{abstract}
  We show that all Majumdar--Papapetrou electrovacuum space--times
with a non--empty black hole region and with a non--singular domain of
outer communications are the standard Majumdar--Papapetrou
space--times.
\end{abstract}

\section{Introduction}
\label{introduction}
Consider an electrovacuum space--time with a non--empty black hole region
$\cal B$
and with an asymptotically flat spacelike surface $\Sigma$ such that
$\partial \Sigma$ is a compact manifold lying in the black hole region.
Suppose further that $|Q|=M$, where $Q$ is the total electric charge
as seen from the asymptotically flat region of $\Sigma$ and $M$ is the
ADM mass of $\Sigma$. According to \cite{GibbonsHull,GHHP,Tod}, (under perhaps
some supplementary conditions on $\partial\Sigma$) one expects that
\begin{enumerate}
\item
On $\Sigma$ there is a globally defined Killing vector field $X$ which
is timelike in the asymptotically flat region.
\item\label{condi2}
For any $p\in\Sigma$ such that $X$ is timelike there exists a
neighbourhood ${\cal O}_p$ thereof and a coordinate system
$x^\mu\in\Omega_p\subset\R^4$ such that the gravitational and
electromagnetic fields take the Israel--Wilson--Perjes
\cite{IW,Perjes} form.
\item
The ADM four--momentum of $\Sigma$ is timelike.
\end{enumerate}
This leads naturally to the question of classifying space--times with
the above properties. To our knowledge no conclusive study of this
problem has been done so far ({\em cf.}, however, \cite{HartleHawking} for
some remarks related to this issue).
In this letter we wish to settle this
question under the supplementary assumption that the domain of outer
communications is static, {\em i.e.}, that the twist of the Killing
vector field vanishes. In that case in the local coordinates discussed
above the metric $g$ and the electromagnetic potential $A$ can be
written in the Majumdar--Papapetrou (MP) form
\cite{Majumdar,Papapetrou}
\beqa\label{I.0}
& g = -u^{-2}dt^2 + u^2(dx^2+dy^2+dz^2)\,, &
\\
\label{I.0.1}
 &A = u^{-1}dt\,, &
\eeqa
with some nowhere vanishing, say positive, function
$u$. Einstein--Maxwell equations read then
\beq
\label{I.1}
\frac{\partial u}{\partial t}=0\,, \qquad
\frac{\partial^2u}{\partial x^2} +
\frac{\partial^2u}{\partial y^2} +
\frac{\partial^2u}{\partial z^2} = 0 \,.
\eeq
A space--time will be called a standard MP space--time if the
coordinates $x^\mu$ of \eq{I.0}--\eq{I.0.1} cover the range
$\R\times(\R^3\setminus\{\vec a_i\})$ for a finite set of points $\vec
a_i\in\R^3$, $i=1,\ldots,I$, and if the function $u$ has the form
\beq
\label{standard}
u=1+\sum_{i=1}^I \frac{m_i}{|\vec x - \vec a_i|} \,,
\eeq
for some positive constants $m_i$. It has been shown by Hartle and
Hawking \cite{HartleHawking} that every standard MP space--time can be
analytically extended to an electro--vacuum space--time with a
non--empty black hole region, and with a domain of outer communication
which is non--singular in the sense described below.\footnote{The
case in which
$I=\infty$ has been considered in \cite[Appendix B]{Chnohair}, where
it was pointed out that the scalar $F_{\mu\nu}F^{\mu\nu}$ is unbounded in
such space--times if the $\vec a_i$'s have accumulation points. It
follows from our analysis below that the case where $I=\infty$ and the
$\vec a_i$'s do not have accumulation points cannot lead to regular
asymptotically flat space--times in the sense of Theorem \ref{T1}.
}
We shall prove the following:
\begin{Theorem}
\label{T1}
Consider an electro--vacuum space--time $(M,g)$ with a
non--empty black hole region $\cal B$.
Suppose that there exists in $M$ an asymptotically flat
spacelike hypersurface $\Sigma$ with compact interior, 
with
boundary $\partial\Sigma\subset {\cal B}$ and with timelike (non--vanishing)
ADM four--momentum. Assume moreover that
 on the closure of the domain of outer communication $\doc$ there
exists a Killing vector field $X$ with complete orbits diffeomorphic to $\R$,
$X$ being timelike in an asymptotic region of $\Sigma$.
If $(M,g)$ is locally a MP space--time in the sense
of point \ref{condi2} above, then there exists  a subset of $\doc$
which is isometrically
diffeomorphic to a standard MP space--time.
\end{Theorem}

It is clear that all the hypotheses above are necessary
in the sense that they are satisfied by the standard MP space--times.
In section \ref{further} we present another version of Theorem
\ref{T1}, and we discuss various ways of modifying the hypotheses above.

One would like to strengthen the conclusion of Theorem \ref{T1}
to conclude that
$\doc$ must be isometrically diffeomorphic to a standard MP space--time.
To do that one would need to prove that there are no other extensions
of a standard MP space--time than those constructed by Hartle and Hawking in
\cite{HartleHawking}. This seems to be a difficult
problem, the resolution of  which lies outside the scope of this paper.

\section{Definitions and proof}
\label{proof}
Before passing to the proof of Theorem \ref{T1} we wish to give a few
definitions and to make some preliminary remarks. Let $\Sigma $ be a
spacelike surface in an electro--vacuum space--time $(M,g)$. A set
$\Sigma_{\ext}\subset \Sigma$ will be said to be an {\em asymptotically
flat three--end} if $\Sigma_{\ext}$ is diffeomorphic to
$\R^3\setminus B(R)$, where $B(R)$ is a closed ball of radius $R$ in
$\R^3$. Moreover we shall ask that in the coordinates induced on
$\Sext$ by this identification we have
\beqa &
|g_{ij}-\delta_{ij}|
+|r\partial_k g_{ij}| +|rK_{ij}|+|A_\mu|+|rF_{\mu\nu}| \le Cr^{-\epsilon}\,,
& \\ &
\forall X^i \in \R^3 \qquad C^{-1}\sum (X^i)^2\le g_{ij}X^iX^j\le
C \sum (X^i)^2
\,,
&
\eeqa for
some constant $C$ and some $\epsilon >0$. Here $K_{ij}$ is the
extrinsic curvature tensor of $\Sext$. Finally we require that the
Killing vector be timelike on $\Sext$. A spacelike hypersurface
$\Sigma$ will be said to have {\em compact interior} if there exists a
manifold $\Sint$, the closure of which
is a compact
manifold with boundary, such that $\Sigma=\Sint\cup_{i=1}^I\Sext{}_{,i}$, for
some finite number of asymptotically flat ends $\Sext{}_{,i}$. Moreover
for each $i$ the boundary $\partial\Sext{}_{,i}$ and some
connected component of $\partial\Sint$ are
assumed to be identified by a diffeomorphism.

Let us mention that if $\doc$ is globally
hyperbolic, then Proposition 2.1 of \cite{ChWtopo} shows that there is
no loss of generality in assuming that there is only one asymptotic
end. We shall however not make the assumption
of global hyperbolicity of $\doc$.

Let us from now on choose one of the asymptotically flat ends, and to minimize
notation let us use the symbol $\Sext$  for the end in question.
Consider an electro--vacuum space--time with an asymptotically flat
end $\Sext$ with timelike ADM four--momentum
and with a Killing vector $X$ which is timelike on
$\Sext$. It follows from \cite{ChBeig} that there exists $\epsilon >0$
such that $X^\mu X_\mu < -\epsilon$ for all $r\ge R_1$ for some $R_1$.
(We use the signature $(-,+,+,+)$.)
If the orbits of $X$ through $\Sext$ are complete then by
\cite{Chnohair,KOM,Simon,DamourSchmidt} there exists a conformal completion of
$M$
satisfying the usual completeness requirements \cite{GerochHorowitz}.
We can then define a
black hole region in the standard way \cite{HE} as ${\cal B}=M\setminus
J^-(\scri^+)$, a white hole region as ${\cal W}=M\setminus
J^+(\scri^-)$, and the domain of outer communications
as $\doc=M\setminus({\cal B}\cup{\cal W})$. These definitions coincide then
with those used in \cite{ChWmax}.

Let us now pass to the proof of Theorem \ref{T1}.
Consider the  set
\beq \label{eqT1}
\tilde\Sigma=\{p\in\Sigma:X(p)\ \mbox{ is
timelike}\}\,.
\eeq
If $\tilde\Sigma$ is simply connected, let $\check\Sigma=\tilde\Sigma$,
otherwise let $\check\Sigma$ be the universal cover of $\tilde \Sigma$.
Note that if $\check \Sigma\ne\tilde\Sigma$, then $\check\Sigma$ will have more
than one asymptotically flat end. Choose one of those ends and,
by a slight abuse of notation, call it
$\Sext$. Define finally $\hat \Sigma$
to be that connected component of $\check\Sigma$   which contains $\Sext$.
We define $\hat M$ to be $\R\times\hat\Sigma$  with a metric
$\hat g$ defined
uniquely by the requirements  that
\begin{enumerate}
    \item The vector $\partial/\partial t$
    tangent to the $\R$ factor of $\hat
M$ is a Killing vector,
\item on $\hat\Sigma\equiv\{0\}\times\hat\Sigma $ the metric and the extrinsic
curvature coincide with those of the original space--time
\end{enumerate}
({\em cf.\ e.g.\/} \cite[Appendix A, eqs.\ (A.15)--(A.17)]{ChWmax}
for an explicit construction).

On $\tilde\Sigma$ let us define the function $u$ by
\beq \label{uquantity}
u^{-2} \equiv -g_{\mu\nu} X^\mu X^\nu\,.
\eeq
Consider a sequence $p_i\in\tilde\Sigma$ such that $p_i\to p\in
\partial\Sigma$.
By definition of $\tilde\Sigma$ either $p\in\partial\Sigma$ or
$u^{-2}(p_i)\to 0$. In the former case the arguments of \cite[Prop.
3.3]{ChWmax}
show that      $u^{-2}(p_i)\to 0$   as well. Let us by an abuse
of notation denote
by $X$ the Killing vector on $\hat M$, and by $u$ the corresponding
quantity as in \eq{uquantity}. It follows that
\beq
\label{goestozero}
u^{-2}(p)\to_{p\to\partial\hat\Sigma}0 \,.
\eeq

On $\hat M$ we can define an auxiliary metric $h=h_{\mu\nu}dx^\mu
dx^\nu$ by the equation
\beqn
h_{\mu\nu} = u^{-2}(\hat g_{\mu\nu}+u^2X_\mu X_\nu) - u^4 X_\mu X_\nu\,.
\eeqn
By hypothesis around every $p$ in $\hat \Sigma$ there exists a
coordinate system in which $\hat g$ takes the form \eq{I.0}.
It follows  that
$h$ is a flat Lorentzian metric on a neighbourhood of $\hat\Sigma$,
with $X$ being a covariantly constant vector field with respect to
$h$. By isometry invariance this must hold throughout $\hat M$.

Choose a point $p\in\Sext$ and let $\hat e^a$, $a=0,\ldots,3$ be a tetrad
of vector fields at $p$ such that $\hat e^0=X(p) $. $\hat e^a$ should be chosen
orthonormal with respect to the metric $h$. As $\hat M$ is simply
connected and $h$ is flat it follows that the set of equations
\beq
\label{EEq.1}
\hat\nabla_\nu e^{a\mu} =0\,, \qquad  e^{a\mu}(p) = \hat e^{a\mu} \,,
\eeq
admits a unique solution on $\hat M$. Here $\hat\nabla$ is the
Levi--Civita connection of the metric $h$. It then follows from simple
connectedness of $\hat M$ that the set of equations
\beq
\label{EEq.2}
{x^a}_{,\mu}=e^a_\mu\,, \qquad x^\mu(p)=0\,,
\eeq
also admits a unique solution on $\hat M$. The $x^a$'s provide
 a global coordinate system on $\hat M$ in which $g$ takes the
 form \eq{I.0}. It should be clear that the coordinates $x^a$ take
 values in $\R\times(\R^3\setminus\sing)$ for some closed set
 $\sing\subset\R^3$.  When asymptotic flatness is taken into account in
 the above  construction, it is not too difficult to show that $\sing$
 is {\em compact}.

Following \cite{HartleHawking}, we note that
\beq
\label{scalar}
F_{\mu\nu}F^{\mu\nu}= -2 \left((\frac{\partial u^{-1}}{\partial x})^2
+(\frac{\partial u^{-1}}{\partial y})^2
+(\frac{\partial u^{-1}}{\partial y})^2\right)\,.
\eeq
The asymptotic conditions and the interior compactness condition
show that there exists a constant $C$ such that $
F_{\mu\nu}F^{\mu\nu}$ is bounded on $\hat M$, which in turns implies that
\beq
\label{estimate}
|\mbox{grad}\ u^{-1} |\le C_1
\eeq
for some constant $C_1$. Here the norm of the gradient refers to the
flat metric on $\R^3$. Clearly,
$u^{-1}$ is uniformly Lipschitz continuous on $\R^3\setminus \sing$.

We now claim that $\sing $ must be a finite set of points.
In fact, for any $R>0$  we must have
\beq
\label{estimateofpoints}
\#(\sing\cap B(R))< C_1 R u(0) +1 \,.
\eeq
Here $C_1$ is the constant of \eq{estimate}; note that $0\not\in\sing$
by construction so that $u(0)$ is well defined. To prove
\eq{estimateofpoints}, let $N$ be the smallest integer larger than
or equal to
$C_1R u(0)+1$ and suppose that there exist points
$x_1,\ldots,x_N\in\partial\Sigma\cap B(R)$.  Set
\beq\label{rhodef}
\rho = \inf_{i\ne j}|x_i-x_j|\,.
\eeq
Choose any $\delta\in(0,1)$ and consider the function
$$
v(x)=C_1^{-1}(1-\delta)\sum_{i=1}^N\frac{1}{|x-x_i|}
\,.
$$
Let $\sing_\epsilon$ denote an $\epsilon$--thickening of $\sing$ and
let $x\in \partial\sing_\epsilon$. Let $x_1$ be the point closest
to $x$, we then have $|x-x_1|\ge\epsilon$. Consider now a ball
$B_{\rho/2,x}$ of radius $\rho/2$ centered at $x$, with $\rho$ defined
in \eq{rhodef}. If $x_1\in B_{\rho/2,x}$, then no other point $x_i$
can also be in $B_{\rho/2,x}$, hence for $i\ne 1$ we must have
$|x_i-x|\ge\rho/2$. If $x_1\not\in B_{\rho/2,x}$, we must also have
$|x_i-x|\ge\rho/2$ for $i\ne 1$ as $x_1$ was the closest point.
This gives  the estimate
\beqn
v\Big|_{\partial\sing_\epsilon}<\frac{1-\delta}{C_1\epsilon}+
2\frac{N}{C_1\rho}\,.
\eeqn
By \eq{goestozero} the function $u^{-1}$ vanishes
on $\partial\sing$, and  the estimate \eq{estimate}
shows that
\beqn
u^{-1}(x)\le C_1 d(x,\partial\sing)\,,
\eeqn
where $d(x,\partial\sing)$ denotes the distance from $x$ to
$\partial\sing$.
It follows that
\beqn
u\Big|_{\partial\sing_\epsilon}\ge\frac{1}{C_1\epsilon}\,.
\eeqn
We thus have, for all $\delta>0$  and $\epsilon\le\epsilon_0(\delta)$
for some $\epsilon_0(\delta)$,
\beq
\label{1}
(u-v)\Big|_{\partial\sing_\epsilon}>0\,.
\eeq
On the other hand for large $r$ the function $v$ tends to zero while
$u$ tends to $1$ by asymptotic flatness, in fact $u>1$ by the maximum
principle.  Hence we also have that $(u-v)(x)$ is positive for $r(x)$
large enough. Both $u$ and $v$ are harmonic on
$\R^3\setminus\sing_\epsilon$ and thus, by the maximum principle, we
must have $u-v>0$ on $\R^3\setminus\sing_\epsilon$.

Consider now $v(0)$; we clearly have
\beqn
v(0)\ge\frac{N(1-\delta)}{C_1R}\,.
\eeqn
Since $\delta>0$ can be chosen arbitrarily small we conclude that
$$u(0)\ge \frac{N}{C_1R}\,,$$
that is,
$$
u(0)\ge \frac{C_1Ru(0)+1}{C_1R}>u(0)\,,
$$
which gives a contradiction, and \eq{estimateofpoints} follows.

As $\sing$ is compact by construction, we can choose $R$ to be large
enough so that $\sing\cap B(R )=\sing$. This shows that $\sing$ must
be a finite set of points, as claimed. It is now a standard result of
potential theory that $u$ has the form \eq{standard}.

One of the consequences of what has been shown is that $\hat M$
has only one asymptotically flat region. Now if the set $\tilde\Sigma$
defined by \eq{eqT1} had been non--simply connected, then $\hat M$ would have
had more than one such region. We conclude that $\tilde \Sigma$ is
simply connected. This together with the assumed properties
of Killing orbits of $X$ on
$\doc\subset M$ allows us to identify $\hat M$ with a subset of $M$
in the obvious way, and Theorem \ref{T1} follows.

\section{Some alternative results}
\label{further}

Let us start by pointing
out that in Theorem \ref{T1} the hypothesis of existence
of the spacelike surface $\Sigma$ can be replaced by the requirement
that there exists a Cauchy surface for $\doc$ which is a complete
Riemannian manifold with respect to the induced metric, and which has
at least one asymptotically flat end. Note that such
a Cauchy surface will not be asymptotically flat with compact
interior, rather it will have some number of asymptotic ends in which
the metric is not asymptotically flat.

It might be desirable for some purposes to have a formulation of
the result at hand in which no mention of a black hole is made.
First note, that the discussion of \cite{GibbonsHull,GHHP}
can be carried through in a purely three--dimensional context, in which no
global
properties of the resulting developments need to be assumed.
In this way one avoids the rather difficult question of existence
of a development with a sufficiently regular conformal completion.
Moreover if the resulting space--time is not censored, then the definition
of a black hole using $\scri$ might be meaningless. Next, after having assumed
that
there exists a Killing vector field on $\doc$ one would still need to establish
completeness of the orbits thereof (the completeness of the Killing orbits
 does not
{\em a fortiori} follow from
the results of \cite{Chorbits} under the hypotheses
made here).
All these issues can be avoided in a Cauchy data setting if one is willing
to replace the condition that $\partial\Sigma$ be a subset of the black hole
region
by the requirement that $X$ becomes null, or perhaps vanishes,
 on $\partial\Sigma$.
More precisely, we have the following:

                                    \begin{Theorem}
\label{T2}
Consider
an electro--vacuum space--time $(M,g)$ and
suppose that there exists in $M$ an asymptotically flat
spacelike hypersurface $\Sigma$ with compact interior,
with  non--empty
boundary $\partial\Sigma$ and with timelike (non--vanishing)
ADM four--momentum. Assume moreover that
  there
exists a Killing vector field $X$ defined in a neighbourhood of $\Sigma$,
$X$ being timelike in an asymptotic region of $\Sigma$ and null (perhaps
vanishing)
on $\partial\Sigma$.
If $(M,g)$ is locally a MP space--time in the sense
of point \ref{condi2} of Section \ref{introduction},
then there exists  a neighbourhood of $\hat\Sigma$
which is isometrically
diffeomorphic to a subset of a standard MP space--time.   Here
   $\hat\Sigma$    is defined as that connected component of
   $\{p\in\Sigma: X^\mu$ is timelike$\}$ which intersects
   the asymptotically flat region.
\end{Theorem}

The proof of Theorem \ref{T2} is a somewhat simpler version
of the proof of Theorem \ref{T1}.

\section{Closing remarks}

Recall now that while static electrovacuum black holes with
non--degenerate horizons are well understood ({\em cf.\/}
\cite{Heusler} and references therein), those which contain
degenerate horizons have so far elluded any attempts for systematic
classification. We hope that the results of
\cite{GibbonsHull,GHHP,Tod} together with our paper provide a step in this
direction. A complete classification could be achieved if one could
prove that the existence of some component of the horizon which is
degenerate implies $M=|Q|$. Unfortunately, it seems that even in the
case of a connected degenerate horizon in a static electro--vacuum
black hole space--time such an equality has not been established so far.

  {\bf Acknowledgements} P.T.C. is grateful to the Schr\"odinger
Institute in Vienna for hospitality during work on this paper.

\end{document}